# Teleportation of the one-qubit state in decoherence environments

Ming-Liang Hu[*]

*School of Science, Xi'an University of Posts and Telecommunications, Xi'an 710061, China*

We study standard quantum teleportation of one-qubit state for the situation in which the channel is subject to decoherence, and where the evolution of the channel state is ruled by a master equation in the Lindblad form. A detailed calculation reveals that the quality of teleportation is determined by both the entanglement and the purity of the channel state, and only the optimal matching of them ensures the highest fidelity of standard quantum teleportation. Also our results demonstrated that the decoherence induces distortion of the Bloch sphere for the output state with different rates in different directions, which implies that different input states will be teleported with different fidelities.

## 1. Introduction

Quantum teleportation, which uses prior shared entanglement between the sender (traditionally named Alice) and the remote receiver (named Bob) as resource, is arguably the most fascinating features and potential applications of quantum mechanics in quantum information science [1]. This protocol enables the recreation of an arbitrary unknown quantum state at a remote place with the help of local quantum operations (i.e., Bell measurements, Pauli rotations, et al.) and classical communication and without the need of transferring any particles physically.

Since the seminal work of Bennett et al. [1], teleportation of single-body quantum state via single quantum channel has received extent investigations both theoretically and experimentally (see, e.g., Refs. [2,3], and references therein). Moreover, teleportation of a quantum state using the mixed entangled state as resource has been theoretically studied recently [4]. Also teleportation of an entangled two-body pure state by using two copies of Werner states as noisy quantum channels was considered by Lee and Kim [5], and they found that the entanglement of the two-qubit state is lost during the teleportation process even when the channel has nonzero quantum entanglement. Besides these, it is shown in Refs. [6,7] that standard teleportation with an arbitrary entangled mixed state resource is equivalent to a generalized depolarizing channel with probabilities given by the maximally entangled components of the resource.

Of the many schemes proposed for teleportation, the scheme for teleporting the one-qubit state using the thermally equilibrium state generated by Heisenberg interactions in solid state systems has been theoretically demonstrated in Refs. [8–13]. In addition, entanglement teleportation (as the name implies, entanglement teleportation means teleporting an entangled quantum state) of various two-qubit states through two independent Heisenberg chains have also been investigated recently [14–16]. These investigations reveal several interesting aspects of quantum entanglement not reflected by the concurrence [17] of the state. Particularly, sometimes a more entangled state can even not be able to ensure quantum teleportation with fidelity (see Section 2) better than a less entangled one, this leaves open the question what is the underlying physical essence behind these counterintuitive phenomena, or in other words, which quantity determines quality of the teleported state except entanglement of the resource?

---

[*] *E-mail*: mingliang0301@xupt.edu.cn, mingliang0301@163.com (M.-L. Hu)

Although the previous results showing that the thermally mixed entangled states can enable quantum teleportation with fidelity better than that obtained only by classical communication, the practical implementation of these protocols has to face the facts that real quantum systems can never be perfectly isolated from the surrounding world, which may impacts on the fidelity of the expected outcomes [5,6]. Indeed, Oh et al. have demonstrated that the type of noises acting on the quantum channels determines the range of states that can be accurately teleported [18]. Moreover, Jung et al. examined quantum teleportation of one-qubit state with a noisy EPR state and showed analytically that all entanglement measures decay to zero when the average fidelity is smaller than $2/3$, and thus the entanglement can be regarded as a genuine resource for teleportation [19]. Also Jung et al. examined two-party quantum teleportation via mixed state quantum channels generated by different noises acting on the initial Greenberger-Horne-Zeilinger (GHZ) and *W* states [20], through which they demonstrated explicitly that the robustness between GHZ and *W* states in quantum teleportation depends on the noisy types. Moreover, teleportation with the common spin environment acting on the qubits in possession of Alice has also been theoretically studied [21].

Motivated by the above works, in this paper we present our results on standard teleportation of one-qubit state via the mixed state quantum channel generated from the two-qubit Heisenberg *XY* chain. In contrast to the bulk of the previous studies, we consider the case that the channel state subjects to decoherence and its evolution is ruled by a master equation in the Lindblad form [22]. Although the model we considered is somewhat similar to that in Ref. [18], we concentrated on, however, different system-environment coupling mechanisms (see Section 2 for more detail). We find that the larger amount of quantum entanglement does not always yield the better teleportation fidelity, in accordance with Ref. [9]. In fact, as indicated in the following text, the quality of the teleported state is determined by both the entanglement and the purity of the channel state, and only the optimal matching of these two quantities ensures the highest fidelity of standard quantum teleportation.

## 2. The Formalism

In the present paper, we examine efficiency of quantum teleportation of one-qubit state via the standard protocol $\mathcal{P}_0$ [1], and with the addition of the anisotropic two-qubit system (serves as the quantum channel) subject to decohering environment, and governed by the flowing Heisenberg *XY* Hamiltonian

$$\hat{H} = \frac{J+\Delta}{2}\sigma_A^1\sigma_B^1 + \frac{J-\Delta}{2}\sigma_A^2\sigma_B^2, \tag{1}$$

where $(J+\Delta)/2$ and $(J-\Delta)/2$ are the coupling strengths in the *x* and *y* directions, respectively, with the parameter $\Delta$ measures the anisotropy of the system in the *x*-*y* plane and equals to 0 for the isotropic *XX* model and $\pm J$ for the Ising model. $\sigma_\alpha^i$ ($i=1,2,3$) signify the usual Pauli operators at site $\alpha = A, B$.

The main purpose of this work is to examine the physical role of decoherence in the process of quantum teleportation. In order to address this issue it is convenient to assume that each individual spin of the system that constitute the quantum channel interacts independently with the decohering environment, then under the assumption of Markov and Born approximations and after performing the partial trace over the environmental degrees of freedom, the dynamical behaviors of this open quantum system can be described by the following master equation in the Lindblad form [22]

$$\frac{d\rho^c}{dt} = -i[\hat{H}, \rho^c] + \frac{\gamma}{2}\sum_\alpha \mathcal{L}_\alpha \rho^c, \tag{2}$$

where $\rho^c$ and $\gamma$ denote, respectively, the density matrix of the system and the coupling strengths of the qubits with their respective environments (i.e., we assume the same system-environment coupling strengths for the two qubits involved in the channel). The Lindblad superoperator $\mathcal{L}_\alpha$ [22] is defined by $\mathcal{L}_\alpha \rho^c = \Sigma_n (2 c_n \rho^c c_n^\dagger - \{c_n^\dagger c_n, \rho^c\})$, with the brace $\{\}$ denotes anticommutator, and in contrast to Ref. [18], the system-environment coupling operators $c_n$ here are expressed in terms of the raising and lowering operators $\sigma_\alpha^\pm = (\sigma_\alpha^1 \pm i\sigma_\alpha^2)/2$ as $c = \sigma_\alpha^-$ for the dissipative environment (i.e., zero temperature reservoir), $c_1 = \sigma_\alpha^-$ and $c_2 = \sigma_\alpha^+$ for the noisy environment (i.e., infinite temperature reservoir), and $c = \sigma_\alpha^+ \sigma_\alpha^-$ for the dephasing environment [23].

Without loss of generality, we consider as input the one-qubit state needs to be teleported to Bob in the form of $|\varphi_{in}\rangle = \cos(\theta/2)|0\rangle + e^{i\phi}\sin(\theta/2)|1\rangle$, where $\theta \in [0, \pi]$ and $\phi \in [0, 2\pi]$ are the polar and azimuthal angles, respectively. Then when the state $\rho^c$ described by Eq. (2) acts as the quantum channel, by adopting the standard teleportation protocol $\mathcal{P}_0$, one can obtain the explicit form of the output state at the receiver Bob's hands as [6]

$$\rho_{out}^{(m)} = \sum_{k=0}^{3} \langle \psi_{\text{Bell}}^{k\oplus m} | \rho^c(t) | \psi_{\text{Bell}}^{k\oplus m} \rangle \sigma^k \rho_{in} \sigma^k . \tag{3}$$

Here $k \oplus m$ ($m = 0,1,2,3$) represents summation modulus 4, and $\rho_{in} = |\varphi_{in}\rangle\langle\varphi_{in}|$. Moreover, $|\psi_{\text{Bell}}^i\rangle$ ($i = 0,1,2,3$) in Eq. (3) denote the familiar four maximally entangled Bell states which are given by $|\psi_{\text{Bell}}^{0,3}\rangle = (|00\rangle \pm |11\rangle)/\sqrt{2}$, and $|\psi_{\text{Bell}}^{1,2}\rangle = (|01\rangle \pm |10\rangle)/\sqrt{2}$. Note that although the expression of Eq. (3) is somewhat different from that of Eq. (20) in Ref. [11], they are essentially the same.

If one defines $\chi_i = \langle \psi_{\text{Bell}}^i | \rho^c(t) | \psi_{\text{Bell}}^i \rangle$, then we obtain

$$\chi_{0,3} = \frac{1}{2}[\rho_{11}^c(t) + \rho_{44}^c(t) \pm \rho_{14}^c(t) \pm \rho_{41}^c(t)],$$
$$\chi_{1,2} = \frac{1}{2}[\rho_{22}^c(t) + \rho_{33}^c(t) \pm \rho_{23}^c(t) \pm \rho_{32}^c(t)], \tag{4}$$

from which Eq. (3) simplifies to

$$\rho_{out}^{(m)} = \sum_{k=0}^{3} \chi_k^{(m)} \sigma^k \rho_{in} \sigma^k , \tag{5}$$

where we have used the notation $\chi_k^{(m)} = \chi_{k\oplus m}$.

To characterize the quality of the teleported state $\rho_{out}^{(m)}$ in decohering environments, it is often quite useful to look at the fidelity [24] between the input state $\rho_{in}$ and the output state $\rho_{out}^{(m)}$, defined as $f^{(m)} = \langle \varphi_{in} | \rho_{out}^{(m)} | \varphi_{in} \rangle$. This quantity gives the information of how close the teleported state $\rho_{out}^{(m)}$ is to the unknown state $\rho_{in}$, i.e., they are equal when $f^{(m)} = 1$ and orthogonal when $f^{(m)} = 0$. For $\rho_{in} = |\varphi_{in}\rangle\langle\varphi_{in}|$, from Eq. (5) one can derive the teleportation fidelity $f^{(m)}$ as $f^{(m)} = \chi_0^{(m)} + \chi_1^{(m)} \sin^2\theta\cos^2\phi + \chi_2^{(m)} \sin^2\theta\sin^2\phi + \chi_3^{(m)}\cos^2\theta$, by combination of which with $F^{(m)} = (1/4\pi)\int f^{(m)} \sin\theta d\theta d\phi$ (with $4\pi$ being the solid angle) one can obtain

$$F^{(m)} = \chi_0^{(m)} + \frac{1}{3}[\chi_1^{(m)} + \chi_2^{(m)} + \chi_3^{(m)}] = \frac{2\chi_0^{(m)} + 1}{3} , \tag{6}$$

where $F^{(m)}$ is the average fidelity (the fidelity $f^{(m)}$ averaged over all possible pure input states $|\varphi_{in}\rangle$ on the Bloch sphere), and we have used the equality $\Sigma_{i=0}^{3} \chi_i^{(m)} = 1$ in deriving the above equation. For a given channel state $\rho^c(t)$, one can implement the standard quantum teleportation protocol $\mathcal{P}_0$ by choosing certain values of $m$ ($m = 0,1,2,3$) to maximize the average fidelity, or in another word, the maximum average fidelity achievable is given by

$$F = \frac{2\mathcal{F}[\rho^c(t)] + 1}{3}, \tag{7}$$

where the fully entangled fraction is expressed as

$$\mathcal{F}[\rho^c(t)] = \max_{m=0,1,2,3} \{\chi_0^{(m)}\} = \max_{m=0,1,2,3} \{\chi_m\}. \tag{8}$$

Eq. (7) gives the maximal average fidelity achievable from $\rho^c(t)$ via the standard teleportation protocol [6, 7]. In order to teleport $|\varphi_{in}\rangle$ with fidelity better than purely classical communication protocol, we require Eq. (7) to be strictly greater than $2/3$ [25].

On the other hand, it is well known that the one-qubit state can also be expressed neatly in the Bloch sphere representation as $\rho = (I + \vec{r} \cdot \vec{\sigma})/2$, where $\vec{r} = (x, y, z)$ is the real three-dimensional Bloch vector for the geometric description of quantum information (when $|\vec{r}| = 1$ the state $\rho$ is pure, and when $|\vec{r}| < 1$ it is mixed). $I$ is the $2 \times 2$ identity matrix, and $\vec{\sigma} = (\sigma^1, \sigma^2, \sigma^3)$ are the familiar Pauli operators. If the explicit form of $\rho$ is known, then the components of the Bloch vector $\vec{r}$ can be expressed in terms of the matrix elements of $\rho$ as $x = 2\operatorname{Re}(\rho_{12}) = 2\operatorname{Re}(\rho_{21})$, $y = -2\operatorname{Im}(\rho_{12}) = 2\operatorname{Im}(\rho_{21})$, and $z = 2\rho_{11} - 1 = 1 - 2\rho_{22}$ [with Re($a$) and Im($a$) represent the real and imaginary part of $a$]. For the special case $\rho = \rho_{in}$ (i.e., the state $\rho$ is pure) the above three expressions simplify to $x_0 = \sin\theta\cos\phi$, $y_0 = \sin\theta\sin\phi$, and $z_0 = \cos\theta$.

If the system is ideally protected from its surrounding environment and any other disturbances, then the single-qubit state $\rho_{in}$ can be perfectly teleported from Alice to Bob by using any one of the four maximally entangled Bell states as the quantum channel [1]. However, in the real circumstance, the quantum channel is inevitably subject to decoherence and decay process, which induces the distortion of the Bloch sphere for $\rho_{out}$, and thus the perfect quantum teleportation of one-qubit state cannot be achieved. For the standard teleportation protocol $\mathcal{P}_0$, the components of the Bloch vector in the three directions can be obtained from Eq. (5) as

$$\begin{aligned}
x &= [\chi_0^{(m)} + \chi_1^{(m)} - \chi_2^{(m)} - \chi_3^{(m)}]x_0, \\
y &= [\chi_0^{(m)} - \chi_1^{(m)} + \chi_2^{(m)} - \chi_3^{(m)}]y_0, \\
z &= [\chi_0^{(m)} - \chi_1^{(m)} - \chi_2^{(m)} + \chi_3^{(m)}]z_0.
\end{aligned} \tag{9}$$

In the present work, we would like to investigate decay dynamics of average fidelity when the standard teleportation protocol $\mathcal{P}_0$ is executed in the presence of dissipative, noisy and dephasing environments [23]. For every type of environment, we analyze the case that both the two qubits involved in the quantum channel subject to decoherence. This consideration is stimulated by the observation that in the real experiments, the entangled state resource used for teleportation may be prepared initially by a third party who then sends one qubit to Alice and the other one to Bob to establish the shared entangled state. During the transmission process, the qubits must be exposed to decohering environments, and degrade quantum entanglement between them.

## 3. Teleportation in dissipative environment

In this section, we would like to explore effects of dissipative environment on average fidelity when the teleportation is performed via the standard protocol $\mathcal{P}_0$ and the two-qubit state $\rho^c(t)$ is used as the resource. For simplicity, we consider the initial channel state to be the so-called X state [26], i.e., the bipartite density matrix $\rho^c(0)$ only contains nonzero elements along the main diagonal and anti-diagonal. For this case, by solving the master equation expressed in Eq. (2) one can obtain all nonzero elements of $\rho^c(t)$ analytically, whose explicit expressions are

$$\rho^c_{11}(t) = ae^{-2\gamma t}, \quad \rho^c_{44}(t) = \rho^c_{11}(t) - [b_2\cos(2\Delta t) + ib_1\sin(2\Delta t)]e^{-\gamma t} + c_2,$$

$$\rho^c_{14,41}(t) = \frac{1}{2}[\varrho^c_{14+41} \pm ib_2\sin(2\Delta t) \pm b_1\cos(2\Delta t)]e^{-\gamma t} \mp i\gamma c_1,$$

$$\rho^c_{22,33}(t) = \frac{1}{2}\{1 - \rho^c_{11}(t) - \rho^c_{44}(t) \pm [\varrho^c_{22-33}\cos(2Jt) + i\varrho^c_{23-32}\sin(2Jt)]e^{-\gamma t}\}, \tag{10}$$

$$\rho^c_{23,32}(t) = \frac{1}{2}[\varrho^c_{23+32} \pm i\varrho^c_{22-33}\sin(2Jt) \pm \varrho^c_{23-32}\cos(2Jt)]e^{-\gamma t},$$

where the corresponding time-independent coefficients $c_i$, $b_i$ ($i = 1, 2$), and the time-dependent coefficient $a$ are as follows

$$c_1 = \frac{\Delta}{4\Delta^2 + \gamma^2}, \quad c_2 = \frac{\gamma^2}{4\Delta^2 + \gamma^2}, \quad b_1 = \varrho^c_{14-41} + 2i\gamma c_1, \quad b_2 = \varrho^c_{11-44} + c_2,$$

$$a = \varrho^c_{11} + [(2i\Delta b_1 - \gamma b_2)\sin(2\Delta t) + (2\Delta b_2 + i\gamma b_1)\cos(2\Delta t)]c_1 e^{\gamma t} \tag{11}$$

$$+ (e^{2\gamma t} - 2b_2 - 1)\Delta c_1 - i\gamma b_1 c_1.$$

In the above two equations the abbreviations $\varrho^c_{ij} = \rho^c_{ij}(t=0)$ and $\varrho^c_{ij \pm kl} = \rho^c_{ij}(t=0) \pm \rho^c_{kl}(t=0)$ are defined for the sake of simplicity of representation (the same notations will be used throughout this paper).

When subject to dissipative environment, our results show that both the concurrence $C$ of the channel state and the average fidelity $F$ of standard quantum teleportation are exactly the same for $\rho^c(0) = |\psi^0_{Bell}\rangle\langle\psi^0_{Bell}|$ and $\rho^c(0) = |\psi^3_{Bell}\rangle\langle\psi^3_{Bell}|$, as well as for $\rho^c(0) = |\psi^1_{Bell}\rangle\langle\psi^1_{Bell}|$ and $\rho^c(0) = |\psi^2_{Bell}\rangle\langle\psi^2_{Bell}|$, the only difference is that the corresponding Bloch sphere for $\rho_{out}$ may shrink along different directions (see the following text for more detail), thus for the ease of representation, we restrict our consideration only to the cases of $\rho^c(0) = |\psi^{0,1}_{Bell}\rangle\langle\psi^{0,1}_{Bell}|$ in the following discussion of standard quantum teleportation. The cases for $\rho^c(0) = |\psi^{2,3}_{Bell}\rangle\langle\psi^{2,3}_{Bell}|$ can be analyzed analogously.

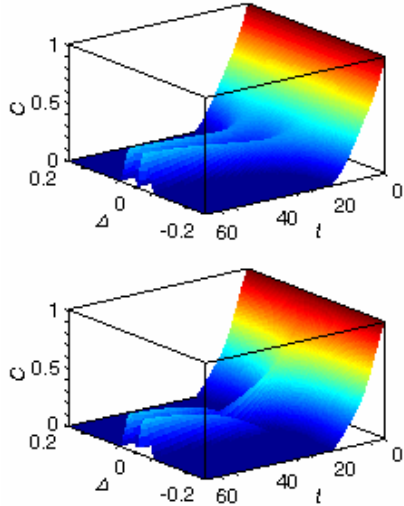

**Fig. 1.** (Color online) Concurrence $C$ of the channel state $\rho^c(t)$ versus the anisotropic parameter $\Delta$ and time $t$ when the system subject to dissipative environment, where the decoherence rate is given by $\gamma = 0.05$. The top and the bottom panels correspond to the case of the channel state $\rho^c(t)$ evolves from $\rho^c(0) = |\psi^0_{Bell}\rangle\langle\psi^0_{Bell}|$ and $\rho^c(0) = |\psi^1_{Bell}\rangle\langle\psi^1_{Bell}|$, respectively.

Before discussing quality of the teleported state, we first see how the dissipative environment affects entanglement dynamics of the channel state $\rho^c(t)$, for entanglement has been considered as an essential resource for teleportation [1,5]. In order to measure quantitatively the amount of entanglement associated with $\rho^c(t)$, we consider the concurrence, a function introduced by Wootters [17], equals to 1 for maximally entangled state and zero for separable state, defined as $C = \max\{0, \lambda_1 - \lambda_2 - \lambda_3 - \lambda_4\}$, where $\lambda_i$ ($i=1,2,3,4$) are the square roots of the eigenvalues of the spin-flipped operator $R = \rho^c(\sigma^2 \otimes \sigma^2)\rho^{c*}(\sigma^2 \otimes \sigma^2)$ in decreasing order of magnitude, with $\rho^{c*}$ being the complex conjugation of $\rho^c$. The concurrence is available no matter whether $\rho^c$ is pure or mixed. For the case of $\rho^c(0) = |\psi^i_{Bell}\rangle\langle\psi^i_{Bell}|$ ($i=1,2,3,4$), it follows from Eq. (10) that the parameter $J$ though appearing in the system Hamiltonian does not contribute to $\rho^c(t)$ because $\varrho^c_{22-33} = \varrho^c_{23-32} = 0$. For this reason, in Fig. 1 we only display plots of the concurrence $C$ versus the anisotropic parameter $\Delta$ and evolution time $t$, where the top and the bottom panels correspond to the case of $\rho^c(t)$ evolves from the initial Bell states $\rho^c(0) = |\psi^0_{Bell}\rangle\langle\psi^0_{Bell}|$ and $\rho^c(0) = |\psi^1_{Bell}\rangle\langle\psi^1_{Bell}|$, respectively, and the decoherence rate is chosen to be $\gamma = 0.05$. One can see that the concurrence is invariant under the substitution $\Delta \to -\Delta$ (one shall see that the same applies to the average fidelity of standard teleportation, the purity of the channel state as well as shrink factors of the output state), and for fixed values of $\Delta$, the entanglement of the channel state $\rho^c(t)$ is suppressed by the dissipative environment with increasing time $t$. Moreover, for $\rho^c(t)$ evolves from $\rho^c(0) = |\psi^0_{Bell}\rangle\langle\psi^0_{Bell}|$, the concurrence may be increased to some extent, though it is not obvious in the short time region from this figure, by properly enlarging the absolute values of $\Delta$. Particularly, when $|\Delta| \gg \gamma$ one can obtain $C = e^{-\gamma t} + (e^{-2\gamma t} - 1)/2$. For $\rho^c(t)$ evolves from $\rho^c(0) = |\psi^1_{Bell}\rangle\langle\psi^1_{Bell}|$, however, the concurrence may be weakened by enlarging the absolute values of $\Delta$ in the short time region (for the parameters chosen in Fig. 1, $t < 33.24$), and when $|\Delta| \gg \gamma$ we also have $C = e^{-\gamma t} + (e^{-2\gamma t} - 1)/2$. In the long time region, as can be seen from the bottom panel of Fig. 1, the concurrence can also be enlarged to some extent by properly increasing the absolute values of $\Delta$.

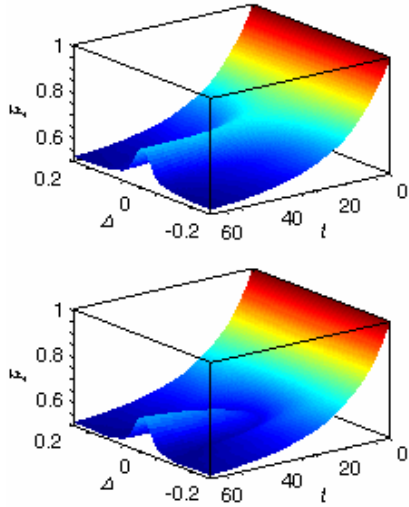

**Fig. 2.** (Color online) Average fidelity $F$ of standard teleportation versus the anisotropic parameter $\Delta$ and time $t$ when the system subject to dissipative environment, where the decoherence rate is given by $\gamma = 0.05$. The top and the bottom panel correspond to the case of the channel state $\rho^c(t)$ evolves from $\rho^c(0) = |\psi^0_{Bell}\rangle\langle\psi^0_{Bell}|$ and $\rho^c(0) = |\psi^1_{Bell}\rangle\langle\psi^1_{Bell}|$, respectively.

In Fig. 2 we show dynamical behaviors of the average fidelity $F$ for the standard teleportation with the decoherence rate also chosen to be $\gamma = 0.05$. For the channel state $\rho^c(t)$ evolves from $\rho^c(0) = |\psi_{Bell}^0\rangle\langle\psi_{Bell}^0|$, one can note that the dissipative environment always makes the average fidelity $F$ decreases with the increase of time $t$ for fixed values of the anisotropic parameter $\Delta$, while for the case of $\rho^c(t)$ evolves from $\rho^c(0) = |\psi_{Bell}^1\rangle\langle\psi_{Bell}^1|$, $F$ decreases with increasing time $t$ from 1 to a certain minimum value that is much smaller than $2/3$ at the critical point $t_c$ ($t_c = \ln 3/\gamma$ if $\Delta = 0$ and increases with the increasing value of $|\Delta|$), and then when $t$ becomes larger than $t_c$, $F$ may be increased to some extent, but cannot exceed the extreme value of $2/3$, which is the maximum achievable only by classical communication [25]. Moreover, the average fidelity $F$ may be increased to some extent in the whole time region by diminishing the absolute values of $\Delta$ for the case that $\rho^c(t)$ evolves from the initial Bell state $\rho^c(0) = |\psi_{Bell}^0\rangle\langle\psi_{Bell}^0|$. For $\rho^c(t)$ evolves from $\rho^c(0) = |\psi_{Bell}^1\rangle\langle\psi_{Bell}^1|$, however, the average fidelity may be increased by enlarging the absolute values of $\Delta$ in the short time region, while in the long time region (for the parameters chosen in Fig. 2, $t > 28.81$), as can be seen from the bottom panel of Fig. 2, the average fidelity of standard teleportation can also be increased to some extent by diminishing the absolute values of $\Delta$, but still cannot exceed the classical limiting value of $2/3$.

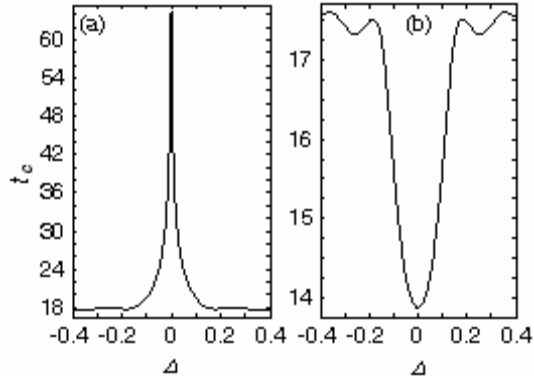

**Fig. 3.** Critical time $t_c$ versus the anisotropic parameter $\Delta$ when the system subject to dissipative environment, and the channel state $\rho^c(t)$ evolves from $\rho^c(0) = |\psi_{Bell}^0\rangle\langle\psi_{Bell}^0|$ (a) and $\rho^c(0) = |\psi_{Bell}^1\rangle\langle\psi_{Bell}^1|$ (b), where the decoherence rate is given by $\gamma = 0.05$. Note that in (a) the critical time $t_c \to \infty$ when $\Delta = 0$.

It is also clear from Fig. 2 that for any fixed values of the anisotropic parameter $\Delta$, there exists a critical time $t_c$ beyond which the performance of quantum teleportation is worse than what purely classical communication protocol can offer [25], i.e., when $t > t_c$ we have $F < 2/3$. Fig. 3 gives the anisotropy dependence of $t_c$ for the cases of the channel state $\rho^c(t)$ evolves from $\rho^c(0) = |\psi_{Bell}^0\rangle\langle\psi_{Bell}^0|$ (the left panel) and $\rho^c(0) = |\psi_{Bell}^1\rangle\langle\psi_{Bell}^1|$ (the right panel), where the decoherence rate is still chosen to be $\gamma = 0.05$. For $\rho^c(t)$ evolves from $\rho^c(0) = |\psi_{Bell}^0\rangle\langle\psi_{Bell}^0|$, one can observe that the critical time $t_c$ can be increased by diminishing the absolute values of $\Delta$. Particularly, when $\Delta = 0$ we have $t_c \to \infty$, which implies that for this special case, the dissipative quantum channel is always superior to its classical counterpart in the whole time region. For the case of $\rho^c(t)$ evolves from $\rho^c(0) = |\psi_{Bell}^1\rangle\langle\psi_{Bell}^1|$, however, we have $t_c = \ln 2/\gamma$ when $\Delta = 0$, and $t_c$ may be increased by properly enlarging the absolute values of $\Delta$. But this increment is still finite, because even for the limiting case of $\Delta \to \infty$ (which may be a difficult experimental task) one can only obtain $t_c = \ln(\sqrt{2}+1)/\gamma$.

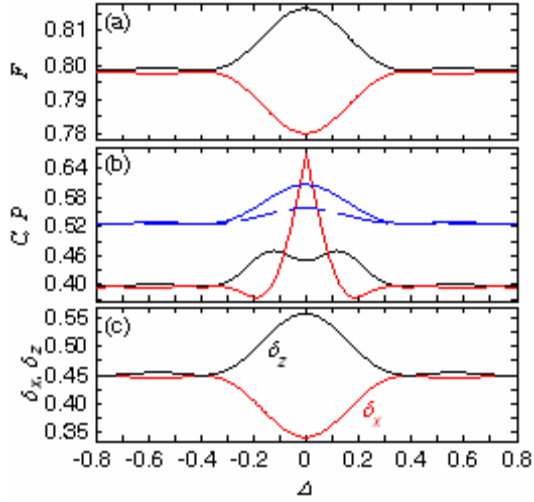

**Fig. 4.** (Color online) Average fidelity of standard teleportation (a), concurrence of the channel state (b) and shrink factors of the Bloch sphere for $\rho_{out}$ versus the anisotropic parameter $\Delta$ when the system subject to dissipative environment. For every plot the black and red curves correspond to $\rho^c(t)$ evolves from $\rho^c(0) = |\psi^0_{Bell}\rangle\langle\psi^0_{Bell}|$ and $\rho^c(0) = |\psi^1_{Bell}\rangle\langle\psi^1_{Bell}|$, respectively. Moreover, the solid and dashed blue curves in (b) show purity of $\rho^c(t)$ versus $\Delta$ with $\rho^c(0) = |\psi^0_{Bell}\rangle\langle\psi^0_{Bell}|$ and $\rho^c(0) = |\psi^1_{Bell}\rangle\langle\psi^1_{Bell}|$, respectively. The other parameters for the plots are given by $\gamma = 0.05$ and $t = 8$.

To show effects of the anisotropy of the system on efficiency of the quantum channel, in Fig. 4 we display average fidelity of standard teleportation, concurrence and purity [27] of the channel state $\rho^c(t)$, as well as shrink factors of the Bloch sphere for $\rho_{out}$ versus $\Delta$ at time $t = 8$ (this choice of time $t$ is based on the consideration that in the region of $t < \ln 2/\gamma$ we always have the average fidelity greater than its classical limiting value of $2/3$), where the shrink factor is defined as $\delta\alpha = |\alpha/\alpha_0|$ ($\alpha = x, y, z$), i.e., the smaller the value of the shrink factor is, the faster the Bloch sphere shrinks. Fig. 4(a) shows again that for the situation in which the channel state $\rho^c(t)$ evolves from $\rho^c(0) = |\psi^0_{Bell}\rangle\langle\psi^0_{Bell}|$, the fidelity of standard teleportation may be increased by diminishing the absolute values of the anisotropic parameter $\Delta$, and when $\Delta = 0$ it attains the maximum $(2+e^{-2\gamma t})/3$. For $\rho^c(t)$ evolves from $\rho^c(0) = |\psi^1_{Bell}\rangle\langle\psi^1_{Bell}|$, however, when $\Delta = 0$ the fidelity $F$ of standard quantum teleportation attains its minimum $(1+2e^{-\gamma t})/3$, and it may be increased to some extent by enlarging the absolute values of $\Delta$. When considering the shrink factors $\delta\alpha$, from Eqs. (4), (9), (10), and (11) one can show that for $\rho^c(0) = |\psi^0_{Bell}\rangle\langle\psi^0_{Bell}|$, the Bloch sphere of $\rho_{out}$ only shrinks or strengths in the $z$ direction when varying the absolute values of $\Delta$ for any fixed time $t$, while for $\rho^c(0) = |\psi^1_{Bell}\rangle\langle\psi^1_{Bell}|$, it only shrinks or strengths in the $x$ direction when varying the absolute values of $\Delta$ for any fixed time $t$. As can be seen from Fig. 4(c), the curves for the shrink factors $\delta z$ and $\delta x$ display similar behaviors as those of the average fidelity shown in Fig. 4(a), which implies that the shrink of the Bloch sphere for $\rho_{out}$ in $z$ and $x$ axis are accompanied by the decrement of the efficiency of standard quantum teleportation. Moreover, by comparing the black and red curves shown in Fig. 4(a) and Fig. 4(b) one can also notice that although the entanglement of the channel state $\rho^c(t)$ for $\rho^c(0) = |\psi^1_{Bell}\rangle\langle\psi^1_{Bell}|$ is larger than that for $\rho^c(0) = |\psi^0_{Bell}\rangle\langle\psi^0_{Bell}|$ in the region of $\Delta \in [-0.083, 0.083]$, however, it does not enable the higher fidelity of standard teleportation. Physically, one could attribute the cause of the poor quality of entanglement to the fact that there is now a comparatively small purity for the channel state $\rho^c(t)$ evolves from $\rho^c(0) = |\psi^1_{Bell}\rangle\langle\psi^1_{Bell}|$ [see the blue curves shown in Fig. 4(b)], which indicates that the concurrence $C$ may only describes certain aspects of the entanglement; it cannot completely reflect the quality of the entanglement as a resource for teleportation. In fact, the success of the standard teleportation protocol $\mathcal{P}_0$ is determined by both the entanglement and the purity of the channel state, and only the optimal matching of these two quantities enables the highest quality of standard teleportation. Same conclusions will be corroborated in the following discussions with other system parameters as well as other scenarios of decohering environments. Also we would like to mention here that a similar conclusion has been demonstrated by Lee and Kim [5], who showed that purity of the initial state determines the possibility of the entanglement teleportation. Here we demonstrated further that purity of the channel state is also very important in determine the quality of the teleported state.

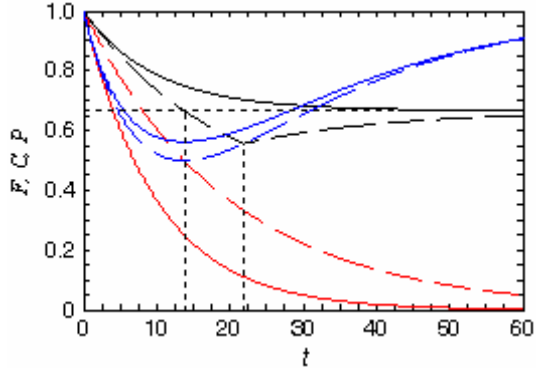

**Fig. 5.** (Color online) Average fidelity of standard teleportation (black curves), concurrence (red curves) and purity (blue curves) of the channel state $\rho^c(t)$ versus time $t$ when the system subject to dissipative environment. For every color the solid and the dashed curves correspond to $\rho^c(0) = |\psi_{Bell}^0\rangle\langle\psi_{Bell}^0|$ and $\rho^c(0) = |\psi_{Bell}^1\rangle\langle\psi_{Bell}^1|$, respectively. The horizontal dotted line at $F = 2/3$ shows the highest fidelity for classical transmission of a quantum state. The other parameters for the plots are given by $\Delta = 0$ and $\gamma = 0.05$.

In the following, we study in detail the relations between average fidelity of standard quantum teleportation of the single-qubit state $|\varphi_{in}\rangle$ via the protocol $\mathcal{P}_0$, entanglement and purity of the channel state $\rho^c(t)$, and shrink factors of $\rho_{out}(t)$ for the special case $\Delta = 0$, which enables an analytical analysis. For $\rho^c(t)$ evolves from $\rho^c(0) = |\psi_{Bell}^0\rangle\langle\psi_{Bell}^0|$, from Eqs. (10) and (11) one can derive its nonzero elements as $\rho_{11}^c(t) = e^{-2\gamma t}/2$, $\rho_{44}^c(t) = 1 - e^{-\gamma t} + e^{-2\gamma t}/2$, $\rho_{14,41}^c(t) = e^{-\gamma t}/2$, and $\rho_{22,33}^c(t) = (e^{-\gamma t} - e^{-2\gamma t})/2$. This, together with Eq. (4) yields

$$\chi_0 = \frac{1}{2}(1 + e^{-2\gamma t}),\ \chi_{1,2} = \frac{1}{2}(e^{-\gamma t} - e^{-2\gamma t}),\ \chi_3 = \frac{1}{2}(1 - 2e^{-\gamma t} + e^{-2\gamma t}). \tag{12}$$

Since $\mathcal{F}[\rho^c(t)] = \max_{m=0,1,2,3}\{\chi_m\}$, one can conclude from Eq. (12) that when $m = 0$ we obtain the maximum average fidelity, which is given by

$$F = \frac{1}{3}(2 + e^{-2\gamma t}). \tag{13}$$

Moreover, the concurrence and purity of the channel state $\rho^c(t)$ at an arbitrary time $t$ can also be obtained analytically as

$$C = e^{-2\gamma t},\ P = 1 - 2e^{-\gamma t} + 3e^{-2\gamma t} - 2e^{-3\gamma t} + e^{-4\gamma t}. \tag{14}$$

From the above two equations, one can see that both the average fidelity of standard quantum teleportation and the pairwise entanglement of the channel state $\rho^c(t)$ decay exponentially with increasing time $t$ (see also the solid lines presented in Fig. 5), and the average fidelity approaches infinitely close to the value $2/3$ while the concurrence goes to zero asymptotically when $t \to \infty$. This indicates that the dissipative quantum channel always enable standard quantum teleportation of one-qubit state with fidelity better than any classical communication channel since $F > 2/3$ in the whole time region. Thus although quantum teleportation does require the quantum channel to be entangled, a nonzero critical value of minimum entanglement is not always necessary, which has been observed previously by Zhou et al. [16] and is in contrast to several previous studies [5, 10,11,14]. Moreover, by combination of Eq. (13) with the first expression of Eq. (14) one can also obtain $F = (2+C)/3$, i.e., when $\rho^c(t)$ evolves from the initial state $\rho^c(0) = |\psi_{Bell}^0\rangle\langle\psi_{Bell}^0|$, the larger the entanglement of the channel state is, the higher the average fidelity of standard quantum teleportation achievable. As for the purity of the channel state $\rho^c(t)$, one can derive from Eq. (14) that $P$ decreases with increasing time $t$ in the region of $t < \ln 2/\gamma$, while out of this region, it begins to increase with the increase of time $t$. However, as can be seen from Fig. 5, the increment of the purity of $\rho^c(t)$ in the region of $t > \ln 2/\gamma$ does not help to enhance the teleportation fidelity, which may be caused by the fact that the entanglement of the channel state $\rho^c(t)$ is weakened with increasing time $t$ during this region. This phenomenon demonstrates again that only the optimal matching of the entanglement and purity of the channel state ensures the highest fidelity of standard quantum teleportation.

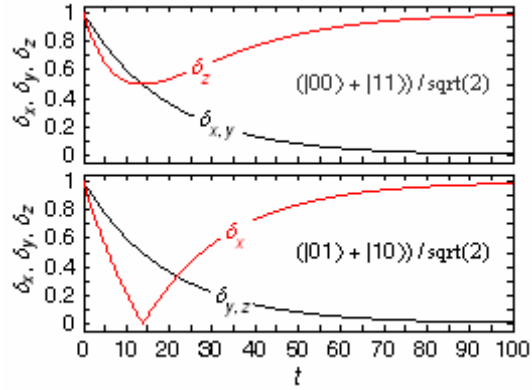

**Fig. 6.** (Color online) The shrink factors $\delta\alpha$ ($\alpha = x, y, z$) versus time $t$ when the system subject to dissipative environment, where the top and the bottom panels correspond to $\rho^c(t)$ evolves from $\rho^c(0) = |\psi_{Bell}^0\rangle\langle\psi_{Bell}^0|$ and $\rho^c(0) = |\psi_{Bell}^1\rangle\langle\psi_{Bell}^1|$, respectively. The other parameters for the plots are given by $\Delta = 0$ and $\gamma = 0.05$.

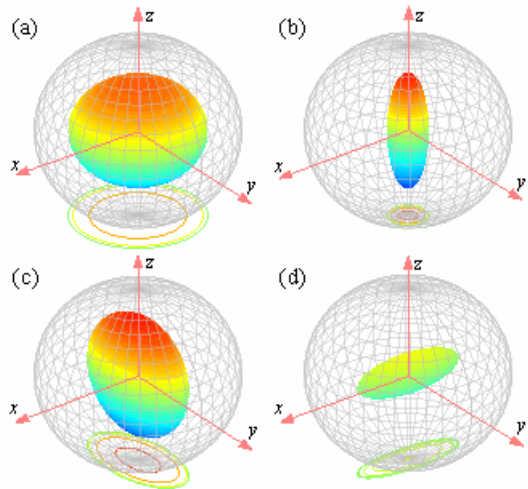

**Fig. 7.** (Color online) The distortion of the Bloch sphere when the system subject to dissipative environment. The gray transparent plots show the Bloch spheres for the initial state $\rho_{in}$, while the chromatic opaque plots show the distorted Bloch spheres for $\rho_{out}(t)$. (a) $\rho^c(0) = |\psi_{Bell}^0\rangle\langle\psi_{Bell}^0|$, $t = 8$; (b) $\rho^c(0) = |\psi_{Bell}^0\rangle\langle\psi_{Bell}^0|$, $t = 32$; (c) $\rho^c(0) = |\psi_{Bell}^1\rangle\langle\psi_{Bell}^1|$, $t = 8$; (d) $\rho^c(0) = |\psi_{Bell}^1\rangle\langle\psi_{Bell}^1|$, $t = 32$. The other parameters for the plots are given by $\Delta = 0$ and $\gamma = 0.05$.

Now we analyze how the dissipative environment affects shrink rates of the Bloch sphere for $\rho_{out}(t)$ with increasing time $t$ for the case of $\rho^c(t)$ evolves from $\rho^c(0) = |\psi_{Bell}^0\rangle\langle\psi_{Bell}^0|$. From Eqs. (9) and (12) one can obtain the three components of the Bloch vector $\vec{r}$ for the output state $\rho_{out}(t)$ as $x = e^{-\gamma t}\sin\theta\cos\phi$, $y = e^{-\gamma t}\sin\theta\sin\phi$, and $z = (1 - 2e^{-\gamma t} + 2e^{-2\gamma t})\cos\theta$, which give rise to $\delta x = \delta y = e^{-\gamma t}$, and $\delta z = 1 - 2e^{-\gamma t} + 2e^{-2\gamma t}$. Clearly, both the shrink factors $\delta x$ and $\delta y$ decay exponentially with increasing time $t$, while $\delta z$ decreases when $t$ increases from $t = 0$ to $t = \ln 2/\gamma$ (at this critical point $\delta z$ attains its minimum $1/2$), and then, it begins to increase with the increase of time $t$ and goes to 1 asymptotically when $t \to \infty$. Moreover, from the top panel of Fig. 6 one can also note that the shrink factors $\delta x$ and $\delta y$ are slightly larger than that of $\delta z$ in the region of $t < \ln 2/\gamma$, which implies that in this region the Bloch sphere shrinks faster in $z$ axis than that in the $x$-$y$ plane. A more intuitive plot is shown in the top two panels of Fig. 7, in which we display the Bloch sphere representation of $\rho_{out}(t)$ for the case of $\rho^c(t)$ evolves from $\rho^c(0) = |\psi_{Bell}^0\rangle\langle\psi_{Bell}^0|$, from which one can also note that when $t < \ln 2/\gamma$ (e.g., $t = 8$), the dissipative environment makes the Bloch sphere shrinks in $x$-$y$ plane uniformly and a little slower than that in $z$ axis. When $t > \ln 2/\gamma$, since $\delta x$ and $\delta y$ continue to decrease while $\delta z$ begins to increase with increasing value of time $t$ (see the top panel of Fig. 6), the Bloch sphere continually shrinks in $x$-$y$ plane and begins to stretch in $z$ axis, which makes it quite like an erect rugby [see Fig. 7(b) for $t = 32$].

For the situation in which the channel state $\rho^c(t)$ evolves from $\rho^c(0) = |\psi_{Bell}^1\rangle\langle\psi_{Bell}^1|$, Eqs. (10) and (11) allow the nonzero elements of $\rho^c(t)$ to be written more explicitly as $\rho_{22,33,23,32}^c(t) = e^{-\gamma t}/2$ and $\rho_{44}^c(t) = 1 - e^{-\gamma t}$. After calculation by substituting this result into Eq. (4) we obtain

$$\chi_{0,3} = \frac{1}{2}(1 - e^{-\gamma t}), \ \chi_1 = e^{-\gamma t}, \ \chi_2 = 0. \tag{15}$$

Different from that of $\rho^c(t)$ evolves from $\rho^c(0) = |\psi_{Bell}^0\rangle\langle\psi_{Bell}^0|$, the relative magnitudes of $\chi_m$ ($m = 0, 1, 2, 3$) in the present case is dependent on the values of time $t$. If $t < \ln 3/\gamma$, we have $\mathcal{F}[\rho^c(t)] = \max_{m=0,1,2,3}\{\chi_m\} = \chi_1$, i.e., when $m = 1$ we obtain the maximum average fidelity. If $t > \ln 3/\gamma$, however, $\mathcal{F}[\rho^c(t)] = \max_{m=0,1,2,3}\{\chi_m\} = \chi_{0,3}$, i.e., now the maximum average fidelity is achieved for both the cases of $m = 0$ and $m = 3$. Thus the average fidelity in the whole time region can be expressed as

$$F = \begin{cases} \dfrac{1 + 2e^{-\gamma t}}{3} & (t \leq \ln 3/\gamma), \\ \dfrac{2 - e^{-\gamma t}}{3} & (t > \ln 3/\gamma), \end{cases} \tag{16}$$

Moreover, the concurrence and purity of the channel state $\rho^c(t)$ are given by

$$C = e^{-\gamma t}, \ P = 1 - 2e^{-\gamma t} + 2e^{-2\gamma t}. \tag{17}$$

It is clear from Eq. (16) that the average fidelity $F$ of standard quantum teleportation decays exponentially when time $t$ increases from $t = 0$ to $t = \ln 3/\gamma$, at which $F$ attains its minimum value $F_{min} = 5/9$, which is slightly smaller than the best possible score $2/3$ achieved by purely classical communication [25]. When $t$ increases after $t > \ln 3/\gamma$, however, $F$ begins to increase monotonously with the increase of time $t$ and approaches to its asymptotic value $2/3$ in the limit of $t \to \infty$ (see the black dashed curve shown in Fig. 5). Moreover, it follows from Eqs. (16) and (17) that in the region of $t < \ln 3/\gamma$ we have $F = (1 + 2C)/3$, while out of this region we have $F = (2 - C)/3$. Therefore in the region of $t > \ln 3 / \gamma$ we obtain again that the large amount of entanglement for the channel state $\rho^c(t)$ does not ensure the high fidelity of standard quantum teleportation, again due to the relative small purity of the channel state (see the blue dashed curve shown in Fig. 5). In Refs. [18,19], the authors argued that the entanglement is a genuine resource for teleportation even if noises are involved, however, as indicated by the dashed curves shown in Fig. 5, this conclusion is not universal because in the present case the entanglement measured by concurrence does not become zero when $F < 2/3$.

When considering the shrink factors for $\rho_{out}(t)$, Eqs. (9) and (15) allow them to be written analytically as $\delta x = |2e^{-\gamma t} - 1|$ and $\delta y = \delta z = e^{-\gamma t}$, where for the case of $t > \ln 3 / \gamma$ we have chosen $m = 3$ (the case for $m = 0$ can be analyzed similarly). Clearly, both $\delta y$ and $\delta z$ decay exponentially in the whole time region, while $\delta x$ initially decreases with increasing time $t$ and attains its minimum value 0 at the critical point $t = \ln 2 / \gamma$, then when $t > \ln 2 / \gamma$, it begins to increase with the increase of time $t$ and approaches to its maximum 1 finally. Moreover, from the plots shown in the bottom panel of Fig. 6 one can also conclude that the dissipative environment makes the Bloch sphere shrinks in $x$ axis faster than that in $y$-$z$ plane over the time region between $t = 0$ and $t = \ln 2 / \gamma$ [see also Fig. 7(c) for $t = 8$], while out of this region, the Bloch sphere continues to shrink in $y$-$z$ plane and begins to stretch in $x$ axis, thus it is quite like an rugby lying on the $x$-$y$ plane [see Fig. 7(d) for $t = 32$].

Before ending this section, we would also like to point out that as indicated by the above discussions, different Bell states generate different Bloch spheres for the teleported state (see Fig. 7), and this does not happen when implementing the standard teleportation protocol through ideal quantum channel, thus depending on the nature of the Bloch sphere, various Bell states distributed by a third party and shared by Alice and Bob may be distinguished.

## 4. Teleportation in noisy and dephasing environment

We now discuss standard teleportation of the one-qubit state with the noisy environment acting on the quantum channel. Similar to the above section, we still consider the channel state (prepared by a third party at $t = 0$ and transmits to Alice and Bob down a noisy channel) $\rho^c(0)$ in an "X" formation [26], for which the density matrix $\rho^c(t)$ can be obtained analytically, which is of the same form as that expressed in Eq. (10), with however the parameter $\gamma$ being substituted with $2\gamma$, and the other parameters $c_1$, $c_2$, and $a$ in Eq. (11) are now given by

$$c_1 = c_2 = 0, \ b_1 = \varrho^c_{14-41}, \ b_2 = \varrho^c_{11-44},$$
$$a = \frac{ib_1 \sin(2\Delta t) + b_2 \cos(2\Delta t)}{2} e^{2\gamma t} + \frac{1}{4} e^{4\gamma t} + \frac{2\varrho^c_{11+44} - 1}{4}. \qquad (18)$$

When $\rho^c(t)$ evolves from $\rho^c(0)$ prepared by a third party in the form of any one of the four maximally entangled Bell states, our calculation shows that they give completely the same results for the average fidelity of standard teleportation, entanglement and purity of the channel state, the only difference is that the corresponding Bloch spheres for $\rho_{out}(t)$ may shrink along different directions. Thus here we restrict our concern only to the case of $\rho^c(0) = |\psi^0_{\text{Bell}}\rangle\langle\psi^0_{\text{Bell}}|$, for which the nonzero elements of $\rho^c(t)$ are given by $\rho^c_{11,44}(t) = (1 + e^{-4\gamma t})/4$, $\rho^c_{14,41}(t) = e^{-2\gamma t}/2$, and $\rho^c_{22,33}(t) = (1 - e^{-4\gamma t})/4$, by substituting of which into Eq. (4) yields

$$\chi_{0,3} = \frac{1}{4}(1 \pm 2e^{-2\gamma t} + e^{-4\gamma t}), \ \chi_{1,2} = \frac{1}{4}(1 - e^{-4\gamma t}). \qquad (19)$$

It follows from Eq. (19) and $\mathcal{F}[\rho^c(t)] = \max_{m=0,1,2,3}\{\chi_m\}$ that for the situation $m = 0$ we obtain the maximum average fidelity, which is given by

$$F = \frac{1}{6}(3 + 2e^{-2\gamma t} + e^{-4\gamma t}). \qquad (20)$$

The concurrence and the purity of the channel state $\rho^c(t)$ can also be obtained analytically as

$$C = \frac{1}{2}\max\{0, 2e^{-2\gamma t} + e^{-4\gamma t} - 1\}, \quad P = \frac{1}{4}(1 + 2e^{-4\gamma t} + e^{-8\gamma t}). \qquad (21)$$

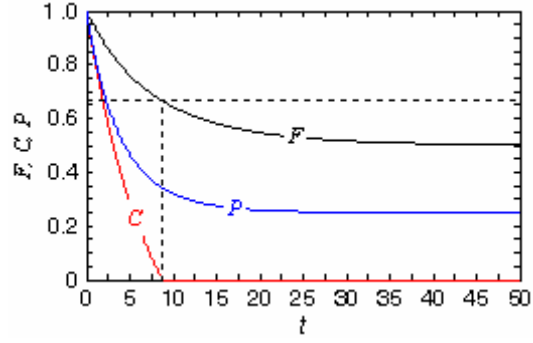

**Fig. 8.** (Color online) Average fidelity of standard teleportation, concurrence and purity of the channel state $\rho^c(t)$ versus time $t$ when the system subject to noisy environment. Here the decoherence rate is given by $\gamma = 0.05$, and the horizontal dotted line at $F = 2/3$ shows the highest fidelity for classical transmission of a quantum state.

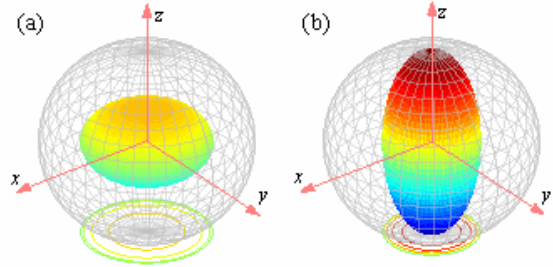

**Fig. 9.** (Color online) The distortion of the Bloch sphere when the system subject to noisy (a) and dephasing (b) environments, where the decoherence rate is given by $\gamma = 0.05$. The gray transparent plots show the Bloch sphere for the initial state $\rho_{in}$, while the chromatic opaque plots show the distorted Bloch sphere for $\rho_{out}(t)$. Moreover, the left panel is plotted with $t = 5$, while the right panel is plotted with $t = 16$.

For the decoherence rate $\gamma = 0.05$, plots of the average fidelity $F$, concurrence $C$ and purity $P$ as functions of time $t$ are shown in Fig. 8. As a first observation, one can find that the noisy environment makes $F$ and $P$ decrease monotonously with increasing time $t$ in the whole time region, while it disentangling the channel state in a finite time $t = \ln(\sqrt{2}+1)/2\gamma$, which is known as entanglement sudden death (ESD) observed previously by Yu and Eberly [28] and has been widely studied in recent years [23,26]. Eqs. (20) and (21) also yield $F = (2+C)/3$ in the region of $t < \ln(\sqrt{2}+1)/2\gamma$, i.e., the larger the amount of entanglement of the channel state $\rho^c(t)$ is, the higher the efficiency of standard quantum teleportation achievable for the special case considered here. When $t > \ln(\sqrt{2}+1)/2\gamma$, since the entanglement of $\rho^c(t)$ remains always zero, the average fidelity of standard teleportation is suppressed by the decay of the purity of $\rho^c(t)$ with increasing time $t$, and in the long time limit $t \to \infty$ the average fidelity $F$ decays to its asymptotic value $1/2$, which corresponds to no-communication between Alice and Bob because this fidelity can be obtained when Bob merely selects a state at random [20].

Moreover, from Eqs. (9) and (19) one can derive the three shrink factors as $\delta x = \delta y = e^{-2\gamma t}$ and $\delta z = e^{-4\gamma t}$. Different from those with the system subject to dissipative environment, here all the shrink factors decay monoexponentially with increasing time $t$, with however, different rates along different directions. To exhibit this more intuitively, we display in the left panel of Fig. 9 the distortion of the Bloch sphere for $\rho_{out}(t)$ with time $t=5$, from which one can observe that the Bloch sphere shrinks faster in $z$ axis than that in $x$-$y$ plane, which makes it quite like a flying saucer lying on the equatorial plane. Furthermore, it follows from $\delta x = \delta y = e^{-2\gamma t}$ and $\delta z = e^{-4\gamma t}$ that the Bloch vector $\vec{r}$ falls to zero in the long time limit $t \to \infty$. This means that the qubit to be teleported is completely depolarized and becomes a totally mixed state in this limit, and thus gives us the average fidelity $F = 1/2$ as depicted in Fig. 8.

Finally, we explore average fidelity of the standard quantum teleportation protocol when it is executed in the presence of dephasing environment. For the initial X-type state [26], if one defines $\kappa = \sqrt{\gamma^2 - 16\Delta^2}/2$ and $\nu = \sqrt{\gamma^2 - 16J^2}/2$, then by solving the appropriate master equation (2), one can obtain the nonzero elements of the channel state $\rho^c(t)$ at an arbitrary time $t$ analytically as

$$\rho^c_{11,44}(t) = \frac{1}{2}(\varrho^c_{11,44} \pm d_1 \varrho^c_{11-44} \pm d_2 \varrho^c_{14-41}),$$

$$\rho^c_{14,41}(t) = \frac{1}{2}(\varrho^c_{14+41}e^{-\gamma t} \pm d_3 \varrho^c_{11-44} \pm d_4 \varrho^c_{14-41}),$$

$$\rho^c_{22,33}(t) = \frac{1}{2}(\varrho^c_{22,33} \pm e_1 \varrho^c_{22-33} \pm e_2 \varrho^c_{23-32}),$$

$$\rho^c_{23,32}(t) = \frac{1}{2}(\varrho^c_{23+32}e^{-\gamma t} \pm e_3 \varrho^c_{22-33} \pm e_4 \varrho^c_{23-32}),$$

(22)

where the time-dependent coefficients $d_i$ and $e_i$ ($i = 1,2,3,4$) are given by

$$d_{1,4} = \frac{2\kappa \cosh(\kappa t) \pm \gamma \sinh(\kappa t)}{2\kappa} e^{-\gamma t/2}, \quad d_{2,3} = \frac{2i\Delta}{\kappa}\sinh(\kappa t)e^{-\gamma t/2},$$

$$e_{1,4} = \frac{2\nu \cosh(\nu t) \pm \gamma \sinh(\nu t)}{2\nu} e^{-\gamma t/2}, \quad e_{2,3} = \frac{2iJ}{\nu}\sinh(\nu t)e^{-\gamma t/2}.$$

(23)

When $\rho^c(0)$ equals to one of the four maximally entangled Bell states, our results show again that the average fidelity of standard teleportation, the entanglement and purity of the channel state $\rho^c(t)$ are completely the same, and the only difference is that the corresponding Bloch spheres may shrink along different directions. Thus in the following discussions we only consider the case of $\rho^c(0) = |\psi^0_{\text{Bell}}\rangle\langle\psi^0_{\text{Bell}}|$, for which the nonzero elements of $\rho^c(t)$ simplifies to $\rho^c_{11,44}(t) = 1/2$ and $\rho^c_{14,41}(t) = e^{-\gamma t}/2$, which yields

$$\chi_{0,3} = \frac{1}{2}(1 \pm e^{-\gamma t}), \quad \chi_{1,2} = 0.$$ (24)

Recall that $\mathcal{F}[\rho^c(t)] = \max_{m=0,1,2,3}\{\chi_m\}$, this, together with Eq. (24) indicates that when $m = 0$ we obtain the maximum average fidelity, which is given by

$$F = \frac{1}{3}(2 + e^{-\gamma t}),$$ (25)

Also the concurrence and the purity of the channel state can be obtained explicitly as

$$C = e^{-\gamma t}, \quad P = \frac{1}{2}(1 + e^{-2\gamma t}).$$ (26)

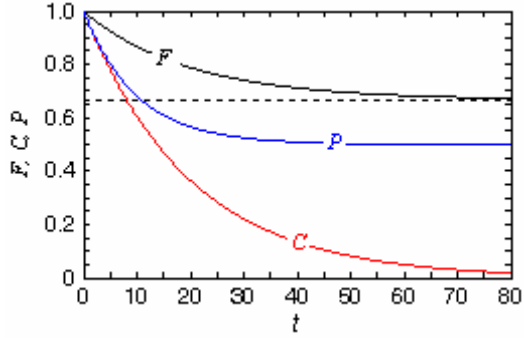

**Fig. 10.** (Color online) Average fidelity of standard teleportation, concurrence and purity of the channel state $\rho^c(t)$ versus time $t$ when the system subject to dephasing environment. Here the decoherence rate is given by $\gamma = 0.05$, and the horizontal dotted line at $F = 2/3$ shows the highest fidelity for classical transmission of a quantum state.

Note the qualitative similarity of the present results for $F$ and $C$ to those of the channel state $\rho^c(t)$ subject to dissipative environment with the addition of $\rho^c(0) = |\psi^0_{Bell}\rangle\langle\psi^0_{Bell}|$ and $\Delta = 0$ (cf. Figs. 5 and 10), this can be easily understood because both $F$ and $C$ here can be obtained by replacement of $2\gamma$ in Eqs. (13) and (14) with $\gamma$. However, the purity of $\rho^c(t)$ here observes different dynamical behaviors for it decays monotonously with increasing time $t$ in the whole time region. Thus the fidelity of standard quantum teleportation is suppressed with increasing time $t$ by the decay of both the entanglement and the purity of the channel state $\rho^c(t)$ when subject to dephasing environment. Moreover, by comparing the present result with that of $\rho^c(t)$ subject to noisy environment, one can also finds that under the condition of the same decoherence rate $\gamma$, the detrimental effects of the noisy environment on teleportation seems to be more severe than that of the dephasing environment.

As for the shrink factors of the teleported state $\rho_{out}(t)$, by inserting Eq. (24) into Eq. (9) one can obtain $\delta x = \delta y = e^{-\gamma t}$ and $\delta z = 1$. Although both $\delta x$ and $\delta y$ here decay exponentially with increasing time $t$, $\delta z$ keeps a constant value of unity, which is dramatically different from all of the former cases considered in the present paper. This indicates that the Bloch sphere for $\rho_{out}(t)$ will shrink in $x$-$y$ plane uniformly with increasing time $t$ and do not shrink in $z$ axis when the system subject to dephasing environment, which makes it quite like an erect rugby (see the right panel of Fig. 9). The above phenomenon also indicates that the input states $|\varphi_{in}\rangle$ near the two poles on the Bloch sphere will be teleported with large fidelities. Particularly, the fidelity is always the maximum 1 at $\theta = \{0, \pi\}$, which correspond to states $|0\rangle$ and $|1\rangle$ that are eigenstates of the system-environment coupling operator $\sigma^+\sigma^-$ describing the dephasing environment.

## 5. Conclusions

In conclusion, by solving analytically the master equation governing the dynamical evolution of the system, we have presented a rather detailed investigation on the ability of the two-qubit spin channel used to teleporting an arbitrary one-qubit pure state via the standard protocol $\mathcal{P}_0$ when being executed in the presence of different decohering environments. By examining the decaying features of the average fidelity of standard teleportation, pairwise entanglement and purity of the channel state, we demonstrated explicitly that the quality of the teleported state is determined by both the entanglement and the purity of the channel state, and only the optimal matching of these two quantities enables the highest fidelity of standard teleportation. The decoherence not only reduces average fidelity of standard quantum teleportation, but also the range of states that can be accurately teleported. Also our results demonstrated that under the influence of the decoherence environments, the corresponding Bloch sphere describing the teleported state at Bob's possession may be distorted in different directions with different rates. This indicates that the different input states will be teleported with different fidelities. Particularly, the shape of the Bloch sphere for the teleported state depends on the explicit forms of the channel states $|\psi_{\text{Bell}}^i\rangle$ ($i = 0,1,2,3$) prepared initially by a third party. Since there is no difference for the Bloch sphere when using any one of the four maximally entangled Bell states as a resource in decoherence-free teleportation process, the decoherence can thus be used to serve as a signature of the type of channel state distributed by a third party.

We also compared the robustness of the four maximally entangled Bell states in terms of their teleportation capacity. In the idealistic situation, they all possess the maximum entanglement and purity and thus enable perfect teleportation of an arbitrary one-qubit state. When being executed in the presence of decohering environments, however, the situation will be quite different. For the case of dissipative environment, the initial states $|\psi_{\text{Bell}}^{0,3}\rangle$ are more robust than $|\psi_{\text{Bell}}^{1,2}\rangle$, while for noisy and dephasing environments, all of them enable teleportation with exactly the same average fidelity. Moreover, the anisotropy of the system also plays an important role in determine quality of the teleportation. For initial states $|\psi_{\text{Bell}}^{1,2}\rangle$ subject to dissipative environment, the fidelity may be enhanced by increasing the absolute values of $\Delta$, however, this increment is finite because it cannot outperform those of the initial states $|\psi_{\text{Bell}}^{0,3}\rangle$. For noisy and dephasing environments with any one of the four Bell states as initial state for the quantum channel, $\Delta$ has no influence on the teleportation fidelity.

Although we concerned in the present work the standard teleportation protocol $\mathcal{P}_0$ with only the entangled state resource subject to decohering environment, still some novel phenomena were obtained. We hope these results may shed some light on further understanding of the general relations between relevant quantities with respect to the quantum channel (e.g., degree of entanglement and magnitude of purity, et al.) and fidelity of quantum teleportation when using spins as qubits.

## Acknowledgments

This work is supported by the NSF of Shaanxi Province under Grant Nos. 2010JM1011 and 2009JQ8006, the Specialized Research Program of Education Department of Shaanxi Provincial Government under Grant No. 2010JK843, and the Youth Foundation of XUPT under Grant No. ZL2010-32.